\documentclass[aps,prl,showpacs]{revtex4}
\usepackage{amsmath}
\usepackage{graphicx}
\usepackage{bm}

\begin{document}

\title{A NEW APPROACH to ANALYSE $H\otimes (2h\oplus g)$ JAHN-TELLER SYSTEM
for $C_{60}$}
\date{\today}
\author{Ramazan Ko\c{c}}
\email{koc@gantep.edu.tr}
\affiliation{Department of Physics, Faculty of Engineering University of Gaziantep, 27310
Gaziantep, Turkey}
\author{Hayriye T\"{u}t\"{u}nc\"{u}ler}
\email{tutunculer@gantep.edu.tr}
\affiliation{Department of Physics, Faculty of Engineering University of Gaziantep, 27310
Gaziantep, Turkey}
\author{Mehmet Koca}
\email{kocam@squ.edu.om}
\affiliation{Department of Physics, College of Science, Sultan Qaboos University, PO Box
36, Al-Khod 123, Muscat, Sultanate of Oman}

\begin{abstract}
It is now well known that electron (hole)-vibron coupling and hence
Jahn-Teller (JT) effect is important understanding the properties of $C_{60}$
and related molecules. In this paper, we study $H\otimes (2h\oplus g)$
coupling case to find the potential energy surfaces for the positively
charged $C_{60}$ molecule due to distortion. The $H\otimes (2h\oplus g)$
Jahn-Teller system is of particular importance as this will be the JT effect
displayed by $C_{60}$ molecules removed with an electron.
$C_{60}^{+}$ is obtained by removing one electron from fivefold degenerate
Hu highest occupied molecular orbital (HOMO) and a hole in HOMO interacts
with the vibrational modes of $C_{60}$ and symmetry is broken. We apply the
method of symmetry breaking mechanism to obtain expressions for the
potential energy surface.
\end{abstract}

\maketitle


\section{INTRODUCTION}

Except at absolute zero in temperature, all atoms and molecules have energy
and vibrate. The electrons in $C_{60}$ molecule will be sensitive to these
vibrations. It is believed that this vibronic coupling plays an important
role in determining the behaviour of $C_{60}$ molecules and related
fullerene compounds.

The symmetry group of $C_{60}$ molecule is icosahedral ($I_{h}$) . This
symmetry group implies large representations, thus large degeneracies of the
interacting electronic and vibrational states of isolated ion. Icosahedral
symmetry is extremely rare in nature, vibronic coupling effects have not
been investigated in this symmetry until very recently. Many interesting
effects due to vibronic coupling are possible from a theoretical point of
view due to quantum-mechanical four and five-fold degenerate states. The
ground state of pure $C_{60}$ molecule is singlet $A_{g}$ state, so it is
not very sensitive to JT effect. However, the ground state of $C_{60}^{+}$
and $C_{60}^{-}$ states do potentially strong effects. Ionic form of $C_{60}$
molecule is subject to the Dynamical Jahn-Teller effect (DJT). The Dynamical
Jahn-Teller effect is closely related to, but should be distinguished from
the static Jahn-Teller effect, where there is a permanent symmetry breaking
molecular distortion or instability to one of the potential-well minima,
removing all or part of orbital degeneracy. JT effect is also important from
experimental point of view. Most workers agree that electron-vibron
interactions are very important for many properties of fullerenes and may
explain superconductivity. It is known that origin of superconducting
pairing in $C_{60}$ compounds is due to interaction of electrons with
vibronic modes of $C_{60}$ molecule.

In the general formalism of the JT effect, a degenerate electronic state
corresponding to a representation of $D$ of symmetry group $G$ of the
molecule can interact with the vibrational modes corresponding to
representations contained in the symmetric part of the direct product $%
D\otimes D$ (excluding the identical representation which is trivial). As it
is well known, the molecular symmetry reduced by the JT distortion with
splitting of the electronic-state degeneracy. The distorting forces acting
along a certain non-totally vibrational modes carry the nuclei over into
distorted configurations. Distorted configuration of a molecule can be
characterised by subgroup symmetries of parent molecular group.

In general, a physical system may pass from a symmetric to less symmetric
state, during its evolution. This symmetry transition is known as
spontaneously symmetry breaking [6,7]. The process of symmetry breaking
applies to all domains described by expectation values of operators. This
common mathematical technique can be applied successfully to different
domains of physics because they all share in the occurrence of broken
symmetry.

In order to study the JT structure of $A_{x}C_{60}$ and/or $C_{60}^{\pm n}$
various theoretical approaches have been used. Ceuleman has proposed on
analytical treatment of JT distortion in the general case of a fivefold
degenerate state of an $I_{h}$ molecule. The extreme points were identified
by using the isostationary function and epikernel methods [1,2]. First order
JT interaction and its continuous group invariance were discussed and the
energies were found for JT coupling with and without splitting [3]. The
ground state Berry phase of some JT systems were calculated in the five
dimensional rotation group SO(5) [4]. Classical and semi-classical models
were introduced about JT manifold for electron vibron interactions, by
Auerback, Manini and Tosatti [5]. The $H\otimes (h\oplus g)$ problem is
investigated analytically using a unitary transformation by C.P Moate, J.L
Dunn et al. [12].

The present paper provides a group theoretical treatment of JT distortion in
general case of a fivefold degenerate state of $I_{h}$ molecule. The JT
surface energies have been obtained by breaking symmetries of $I_{h}$ group
into its maximal little group. This is never done before.

\section{SYMMETRY ASPECTS and COUPLING of STATES}

In this section, we start by describing the Hamiltonian that generates DD
surface where D denotes the irreducible representation. The standard
Hamiltonian may be written in the form%
\begin{equation}
H=H_{0}+H_{JT};\quad H_{0}=\frac{1}{2}\hbar
\sum\limits_{i}(P_{i}^{2}+Q_{i}^{2})  \label{e1}
\end{equation}%
where $H_{0}$ describes free (uncoupled) electrons/holes and vibrations and $%
Q_{i}$ is the distortion coordinate and $P_{i}$ is the conjugate momentum.
In general, $H_{JT}$ introduces a rotationally invariant linear coupling
between the electronic state and the vibrational mode. It is known that the
Hamiltonian for a linear JT coupling of $H\otimes (2h\oplus g)$ system is
invariant under the rotational operations of $I_{h}$ group. If either the
restriction of linear coupling or harmonic forces is relaxed, the symmetry
of Hamiltonian is reduces to point group $G.$

As we have mentioned above, totally symmetric part of direct product of an
irreducible representation of a finite group, which describes the properties
of JT surfaces, is written in the form of%
\begin{equation}
\lbrack D^{\ell }\otimes D^{\ell }]=D^{\ell _{1}}\oplus D^{\ell _{2}}\oplus
\cdots \oplus D^{\ell n}  \label{e2}
\end{equation}%
where is the angular momentum quantum number. Decomposition of $[D^{\ell
}\otimes D^{\ell }]$ implies that the JT Hamiltonian can be written in the
following way%
\begin{equation}
H_{JT}=H^{\ell _{1}}+H^{\ell _{2}}+\cdots +H^{\ell _{n}}  \label{e3}
\end{equation}%
where $H^{\ell }$ is the JT Hamiltonian and it is invariant under the
symmetry operations of corresponding finite group, for the $2\ell +1$
dimensional representation. Since the $I_{h}$ group is subgroup of $O(3)$,
decomposition of coupling of two states should be written in terms of%
\begin{equation}
\lbrack D^{2}\otimes D^{2}]=D^{0}\oplus D^{2}\oplus D^{4}  \label{e4}
\end{equation}%
Therefore, symmetric part of the five dimensional direct product for $I_{h}$
group can be written in the notation of icosahedral molecule:%
\begin{equation}
\lbrack H_{u}\otimes H_{u}]=[H_{u}^{2}]=[H_{g}^{2}]=A_{g}\oplus H_{g}\oplus
(G_{g}\oplus H_{g})  \label{e5}
\end{equation}%
where $A_{g},G_{g}$ and $H_{g}$ are one, four and five dimensional
irreducible representations of $I_{h}$, respectively. The decomposition of
direct product of $[H_{u}\otimes H_{u}]$ coupling in $O(3)$ is found using
the Table 1 and is in the form%
\begin{equation}
H_{JT}=H^{0}+H^{2}+H^{4}  \label{e6}
\end{equation}%
where the superscripts 2 and 4 are $\ell $ values. The symmetric part
contains the totally symmetric representation $H_{0}=A_{g}$ that is trivial
polaronic problem [9,10] and it can be solved, exactly. The $H_{0}$
vibration shift energies but it does not cause splitting. Remaining
vibrations subtend the configuration space, which contains all distorted
configurations, may be reached by JT active coordinates. As shown from Table
1, the Hamiltonian $H_{2}$ corresponds five dimensional representations $%
H_{g}$ and $H_{4}$ corresponds to the direct sum of $G_{g}H_{g}$ vibrational
levels.
\begin{table}[t]
$%
\begin{tabular}{|llll|}
\hline
& $I_{h}$ &  & $I_{h}$ \\ \hline
\multicolumn{1}{|l|}{$0$} & \multicolumn{1}{|l}{$A_{g}$} & $7$ & $%
T_{1u}+T_{2u}+G_{u}+H_{u}$ \\
\multicolumn{1}{|l|}{$1$} & \multicolumn{1}{|l}{$T_{1u}$} & $8$ & $%
T_{2g}+G_{g}+H_{g}+H_{g}$ \\
\multicolumn{1}{|l|}{$2$} & \multicolumn{1}{|l}{$H_{g}$} & $9$ & $%
T_{1u}+T_{2u}+G_{u}+G_{u}+H_{u}$ \\ \cline{2-4}
\multicolumn{1}{|l|}{$3$} & \multicolumn{1}{|l}{$G_{u}+T_{2u}$} & $10$ & $%
A_{g}+T_{1g}+T_{2g}+G_{g}+2H_{g}$ \\
\multicolumn{1}{|l|}{$4$} & \multicolumn{1}{|l}{$G_{g}+H_{g}$} & $11$ & $%
2T_{_{1u}}+T_{2u}+G_{u}+2H_{u}$ \\
\multicolumn{1}{|l|}{$5$} & \multicolumn{1}{|l}{$T_{1u}+T_{2u}+H_{u}$} & $12$
& $A_{g}+T_{1g}+T_{2g}+2G_{g}+2H_{g}$ \\ \hline
$6$ & $A_{g}+T_{1g}+G_{g}+H_{g}$ & $13$ & $T_{1u}+2T_{2u}+2G_{u}+2H_{u}$ \\
\hline
\end{tabular}%
$%
\caption{The representations of the $O(3)$ group with $\ell <$14 are split
by the icosahedral point group.}
\label{tab:b}
\end{table}

\section{DETERMINATION of STATIONARY POINTS on JT SURFACE USING LITTLE
GROUPS of ICOSAHEDRAL}

The second important aspect of invariant polynomial functions, in addition
to, linear Jahn-Teller matrices, concerns the extremum points on JT surface.
Since our interest are the minimal points and the saddle points, the present
section is devoted to the determining of extremum points on JT surface by
breaking symmetries of icosahedral group into its maximal little groups. The
problem has been studied by Ceulemans by using the isostationary function
[1].

The little group is the new symmetry group of distorted molecule as a result
of coupling. A real representation of any subgroup $S<G$; the degree of
subduction can be computed by the relation%
\begin{equation}
C_{\ell }(S)=\frac{1}{S}\sum\limits_{p\in S}\chi _{z}(p)  \label{e7}
\end{equation}%
where $\chi _{z}(p)$ is the character of the representation $p.$In a $z$th
representation of finite group, if all subgroups $S>S$ and $C(S)<C(S)$ then $%
S$ is little group of $G$. For finite groups converse of this definition is
also true. In order to simplify the notation, we shall switch from $O(3)$ to
the $SO(3)$ notation, omitting therefore the $g/u$ symbol for inversion.
Then, the electronic--vibrational coupling is symbolized as $H\otimes
(2h\oplus g)$ where capital letter shows electronic and small letters show
the vibrational levels. Without loss of generality the problem for $I_{h}$
group can be treated equally well in the subgroup of proper rotations of
group $I.$ The maximal little subgroups of $I$group have computed by using
Eqn.(7) and are given in Table 2. As seen from Table 2, the maximal little
groups predict the existence of dihedral groups $D_{5},D_{3}$, and $D_{2}$
minima on the icosahedral molecular cage. In the coordinate space the
distortion that is the result of JT instability should conserve all maximal
little group symmetry.
\begin{table}[t]
$%
\begin{tabular}{|llll|l|}
\hline
& $D_{2}$ & $D_{3}$ & $D_{5}$ & $T$ \\ \hline
\multicolumn{1}{|l|}{$H$} & \multicolumn{1}{|l}{$2A+B_{1}+B_{2}+B_{3}$} & $%
A_{1}+2E$ & $A_{1}+E_{1}+E_{2}$ & $E+T$ \\
\multicolumn{1}{|l|}{$G$} & \multicolumn{1}{|l}{$A+B_{1}+B_{2}+B_{3}$} & $%
A_{1}+A_{2}+E$ & $E_{1}+E_{2}$ & $2A+E$ \\
\multicolumn{1}{|l|}{$H\oplus G$} & \multicolumn{1}{|l}{$%
3A+2B_{1}+2B_{2}+2B_{3}$} & $2A_{1}+A_{2}+3E$ & $A_{1}+2E_{1}+2E_{2}$ & $%
2A+2E+T$ \\ \hline
\end{tabular}%
$%
\caption{Maximal Little groups of $H$ and $G$ representations. Decomposition
implies that $T$ is not maximal little group of $H$ state, and $D_{5}$ is
not maximal little group for $G$ state.}
\end{table}

In our perspective the structure of Jahn-Teller surfaces have been
identified by the symmetry breaking of the continuous symmetry to the true
finite point group of the representation space and its maximal little
groups. This may be represented as follows%
\begin{equation}
I\rightarrow S^{\prime }  \label{e8}
\end{equation}%
where $S^{\prime }$ are little groups of $I$ group given in Table 2.

\section{GROUP INVARIANCE of JAHN-TELLER SYSTEMS}

As we stated in the previous section, the Hamiltonian of $H\otimes (2h\oplus
g)$ coupling have three parts $H^{0},H^{2}$ and $H^{4}$, which must be
separately invariant under $I$ symmetry. In this section, we want to
construct a polynomial function in electronic and nuclear configuration
space, to examine symmetry properties of potential energy surface. A
polynomial function which have been produced for this purpose is in the form%
\begin{equation}
U_{\ell m}(X,Q)=\sum\limits_{i,j=1}^{2\ell
+1}\sum\limits_{k=1}^{2m+1}F_{ijk}X_{i}X_{j}Q_{k}.  \label{e9}
\end{equation}%
In this expression $X_{i}$ and $X_{j}$ correspond to electronic coordinates $%
Q_{k}$ corresponds to nuclear coordinates. The force elements or coupling
coefficients should be chosen appropriately for each $H\otimes h$ and $%
H\otimes (h\oplus g)$ coupling. In the Eqn.(9) the indices $2\ell +1$ and $%
2m+1$ stand for dimensions electronic and active coordinates, respectively.
For $H\otimes h$ coupling $\ell $ and $m$ take value 2 and for $H\otimes
(h\oplus g)$ case$\ell $ and $m$ take values 2 and 4 respectively.

We will focus our treatment for the computation of invariant polynomials for
$H\otimes h$ and $H\otimes (h\oplus g)$ problems. The invariant polynomial
function for $\ell =2$, in real basis has been computed using the matrix
representations of $H$ state which are given in Appendix. The $5\times 5$
generators transform electronic coordinates $X_{i}(i=1,\cdots ,5)$ and
nuclear coordinates $Q_{i}$. It can be written in the form%
\begin{equation}
\sum\limits_{i,j=1}^{5}\sum\limits_{k=1}^{5}F_{ijk}X_{i}X_{j}Q_{k}=\sum%
\limits_{i,j=1}^{5}\sum\limits_{k=1}^{5}F_{ijk}X_{i}^{\prime }X_{j}^{\prime
}Q_{k}^{\prime }.  \label{e10}
\end{equation}%
In this equation, $X_{i}=\sum\limits_{n=1}^{5}\Gamma _{in}^{r}X_{n}$\quad
and\quad $Q_{k}^{\prime }=\sum\limits_{n=1}^{5}\Gamma _{in}^{r}Q_{n}.\quad
\Gamma _{in}^{r}$ is the matrix elements of $5\times 5$ generators of $I$.
The Eqn(10) is solved for $F_{ijk}$ and two linearly independent polynomial
function have been obtained. One of our main goal that first order JT
interaction matrices can be derived by working out invariant polynomial
function. The double differentiation of $U_{22}(X,Q)$ with respect to
electronic coordinates $Xi$ and $X_{j}$ produce linear JT interaction
matrix. In general, we can write%
\begin{equation}
(B_{m})_{ij}=\frac{\partial ^{2}U_{2m}}{\partial X_{i}\partial X_{j}},\quad
(i,j=1,2\cdots 5).  \label{e11}
\end{equation}%
The JT interaction matrix for $H\otimes h$ coupling is%
\begin{eqnarray}
B_{2} &=&F_{1}\left(
\begin{array}{ccccc}
2Q_{5} & 0 & -\sqrt{3}Q_{3} & \sqrt{3}Q_{4} & 2Q_{1} \\
0 & 2Q_{5} & -\sqrt{3}Q_{4} & -\sqrt{3}Q_{3} & 2Q_{2} \\
-\sqrt{3}Q_{3} & -\sqrt{3}Q_{4} & -\sqrt{3}Q_{1}-Q_{5} & -\sqrt{3}Q_{2} &
-Q_{3} \\
\sqrt{3}Q_{4} & -\sqrt{3}Q_{3} & -\sqrt{3}Q_{2} & \sqrt{3}Q_{1}-Q_{5} &
-Q_{4} \\
2Q_{1} & 2Q_{2} & -Q_{3} & -Q_{4} & -2Q_{5}%
\end{array}%
\right) +  \label{e12} \\
&&F_{2}\left(
\begin{array}{ccccc}
\sqrt{15}Q_{1}-\frac{\sqrt{3}}{2}Q_{4}+\frac{\sqrt{5}}{2}Q_{5} & -\sqrt{15}%
Q_{2}+\frac{\sqrt{3}}{2}Q_{3} & \frac{\sqrt{3}}{2}Q_{2} & -\frac{\sqrt{3}}{2}%
Q_{1}-3Q_{5} & \frac{\sqrt{5}}{2}Q_{1}-3Q_{4} \\
-\sqrt{15}Q_{2}+\frac{\sqrt{3}}{2}Q_{3} & -\sqrt{15}Q_{1}+\frac{\sqrt{3}}{2}%
Q_{4}+\frac{\sqrt{5}}{2}Q_{5} & \frac{\sqrt{3}}{2}Q_{1}-3Q_{5} & \frac{\sqrt{%
3}}{2}Q_{2} & \frac{\sqrt{5}}{2}Q_{2}-3Q_{3} \\
\frac{\sqrt{3}}{2}Q_{2} & \frac{\sqrt{3}}{2}Q_{1}-3Q_{5} & -2\sqrt{3}Q_{4}-%
\sqrt{5}Q_{5} & -2\sqrt{3}Q_{3} & -3Q_{2}-\sqrt{5}Q_{3} \\
-\frac{\sqrt{3}}{2}Q_{1}-3Q_{5} & \frac{\sqrt{3}}{2}Q_{2} & -2\sqrt{3}Q_{3}
& 2\sqrt{3}Q_{4}-\sqrt{5}Q_{5} & -3Q_{1}-\sqrt{5}Q_{4} \\
\frac{\sqrt{5}}{2}Q_{1}-3Q_{4} & \frac{\sqrt{5}}{2}Q_{2}-3Q_{3} & -3Q_{2}-%
\sqrt{5}Q_{3} & -3Q_{1}-\sqrt{5}Q_{4} & \sqrt{5}Q_{5}%
\end{array}%
\right)  \notag
\end{eqnarray}%
The force elements $F_{1}$ and $F_{2}$ are given by%
\begin{equation}
F_{1}=\frac{1}{135\sqrt{3}}(-13\sqrt{5}F_{H1}+15F_{H2});\quad F_{2}=\frac{1}{%
27\sqrt{15}}(\sqrt{5}F_{H1}+3F_{H2})  \label{e13}
\end{equation}

In this equation $F_{H1}$ and $F_{H2}$ are coupling parameters of $H$ mode.
The invariant polynomial function is derived for $H\otimes (h\oplus g)$ as
in the same way $H\otimes h$. In this case Eqn.(10) takes form%
\begin{equation}
\sum\limits_{i,j=1}^{5}\sum\limits_{k=1}^{9}F_{ijk}X_{i}X_{j}Q_{k}=\sum%
\limits_{i,j=1}^{5}\sum\limits_{k=1}^{9}F_{ijk}X_{i}^{\prime }X_{j}^{\prime
}Q_{k}^{\prime }.  \label{e14}
\end{equation}%
Nine dimensional $H\otimes (h\oplus g)$ state consists of one $H$ mode with
components $\{Q_{1},Q_{2},Q_{3},Q_{4},Q_{5}\}$ and $G$ mode with components $%
\{Q_{6},Q_{7},Q_{8},Q_{9}\}$. The icosahedral generators transform the
electronic and nuclear coordinates. Transformation $X_{i}$ and $X_{j}$ are
same as given in Eqn.(10). The nuclear coordinates $Q_{k}$ is transformed as
$Q_{k}^{\prime }=\sum\limits_{n=1}^{9}\Lambda _{in}^{\ell }Q_{n},$ where $%
\Lambda ^{\ell }$ is the direct sum of 5 and 4 dimensional irreducible
matrix generators $\Gamma $ and $\Lambda $ given in Appendix, respectively.
Solution of the Eqn.(14) for coefficients $F_{ijk}$ gives that three
linearly independent function. In Ceulemans's paper [1], relations between
polynomial coefficients Fijk in Eqn(10) and Eqn(14) are expressed in terms
of Clebsch Gordon series for the icosahedral point group [13]. The linear JT
interaction matrices for this coupling are derived from the relation (11).
The two of them are same with matrices of $H\otimes h$ coupling and third
one is%
\begin{equation}
F_{3}\left(
\begin{array}{ccccc}
-Q_{6}-2\sqrt{5}Q_{8} & 2\sqrt{5}Q_{7} & \frac{5}{4}Q_{7}-\frac{9\sqrt{5}}{4}%
Q_{9} & -\frac{5\sqrt{5}}{4}Q_{6}-\frac{5}{4}Q_{8} & \frac{\sqrt{15}}{2}Q_{8}
\\
2\sqrt{5}Q_{7} & -Q_{6}+2\sqrt{5}Q_{8} & -\frac{5\sqrt{5}}{4}Q_{6}+\frac{5}{4%
}Q_{8} & \frac{5}{4}Q_{7}+\frac{9\sqrt{5}}{4}Q_{9} & \frac{\sqrt{15}}{2}Q_{7}
\\
\frac{5}{4}Q_{7}-\frac{9\sqrt{5}}{4}Q_{9} & -\frac{5\sqrt{5}}{4}Q_{6}+\frac{5%
}{4}Q_{8} & 4Q_{6}+\sqrt{5}Q_{8} & \sqrt{5}Q_{7} & -\frac{5\sqrt{3}}{2}Q_{7}
\\
-\frac{5\sqrt{5}}{4}Q_{6}-\frac{5}{4}Q_{8} & \frac{5}{4}Q_{7}+\frac{9\sqrt{5}%
}{4}Q_{9} & \sqrt{5}Q_{7} & 4Q_{6}-\sqrt{5}Q_{8} & -\frac{5\sqrt{3}}{2}Q_{8}
\\
\frac{\sqrt{15}}{2}Q_{8} & \frac{\sqrt{15}}{2}Q_{7} & -\frac{5\sqrt{3}}{2}%
Q_{7} & -\frac{5\sqrt{3}}{2}Q_{8} & -6Q_{6}%
\end{array}%
\right) .  \label{e15}
\end{equation}%
The sum of the $B_{2}\ $given in Eqn.(12) and JT interaction matrix given in
Eqn.(15) corresponds to the first order JT interaction matrix ($B_{4}$) for $%
H\otimes (h\oplus g)$ coupling. It is obvious that the interaction matrix is
also obtained by considering only $H\otimes g$ coupling. Force element $%
F_{3} $ is related by coupling parameter of $G$ mode and is given by $%
F_{3}=F_{G}/9.$ The first order linear JT interaction matrices for $H$ state
have been found in [14] and are in agreement with our results. We guess that
the higher order JT interaction matrices may be obtained by constructing
higher order icosahedral invariant polynomials.

In section 4.1, symmetry of icosahedral is broken into its little groups for
$H\otimes h$ and $H\otimes (h\oplus g)$. Combination of eigenvalues of the $%
B_{m}$ with harmonic potential energy, in terms of little groups yield JT
surface energy.

\subsection{TRANSITIONS ASSOCIATED WITH $H\otimes h$ AND $H\otimes (h\oplus
g)$ COUPLINGS}

The five dimensional irreducible representation of I group has three maximal
little group named $D_{2}$, $D_{3}$ and $D_{5}.$ In order to break symmetry
of a parent group into its little groups, one should assign an appropriate $%
Q_{i}$ which can be computed by constructing set of equations such that%
\begin{equation}
Q_{i}=\sum\limits_{j=1}^{2\ell +1}\Gamma _{ij}Q_{j}  \label{e16}
\end{equation}%
where $\Gamma _{ij}$ is the matrix elements of generator of the
corresponding little group. The method of symmetry breaking predicts the
existence of saddle points, trigonal and pentagonal turning points on JT
surfaces associated with $D_{2}$, $D_{3}$ and $D_{5}$ groups. From the
solution of Eqn.(16) it is found that two $H$ type coordinates $%
Q_{H1},Q_{H2} $ and one $G$ type coordinate $Q_{G}.$ The $H$ and $G$ type
coordinates are constrained to $Q_{4}=-\tau Q_{H1},\quad Q_{5}=-\frac{\sqrt{%
15}}{2}Q_{H2},Q_{6}=Q_{G}$, where $\tau =\frac{1}{2}(1+\sqrt{5}).$ Under the
given conditions; the energy eigenvalues of each little group are computed.

\subsubsection{4.1.1 $D_{2}$ TRANSITION}

Decomposition of five dimensional representations in $D_{2}$ group is $%
2A+B_{1}+B_{2}+B_{3}$. The symmetry of the group $I$ is broken into $D_{2}$,
assigning as $Q_{1}\rightarrow \sqrt{5}Q_{4}-\sqrt{3}Q_{5},\quad
Q_{2}\rightarrow 0,\quad Q_{3}\rightarrow 0,\;\;Q_{4}\rightarrow Q_{4},\quad
Q_{5}\rightarrow Q_{5}$, using the Eqn.(16). After substituting values of $%
Q_{i}$ into $B_{2}$, the eigenvalues of $B_{2}$ are carried out.
Combinations of eigenvalues of $B_{2}$ with harmonic restoring potentials
for the distortional coordinates $Q_{H1}$, and $Q_{H2}$ gives JT surface
energy values.%
\begin{eqnarray}
E(A) &=&\pm \frac{1}{2\sqrt{10}}\left[ (3F_{H1}^{2}+F_{H2}^{2})(2\tau
^{2}Q_{H1}^{2}-5\tau Q_{H1}Q_{H2}+5Q_{H2}^{2})\right] ^{\frac{1}{2}}+\frac{1%
}{2}K_{H}(Q_{H1}^{2}+Q_{H2}^{2})  \notag \\
E(B_{1}) &=&\frac{1}{4\sqrt{5}}(F_{H1}+F_{H2})(4\tau
Q_{H1}-5Q_{H2})+(F_{H1}-3F_{H2})\sqrt{5}Q_{H2}+\frac{1}{2}%
K_{H}(Q_{H1}^{2}+Q_{H2}^{2})  \notag \\
E(B_{2}) &=&\frac{1}{4\sqrt{5}}(F_{H2}-F_{H1})(4\tau
Q_{H1}-5Q_{H2})+(F_{H1}+3F_{H2})\sqrt{5}Q_{H2}+\frac{1}{2}%
K_{H}(Q_{H1}^{2}+Q_{H2}^{2})  \notag \\
E(B_{3}) &=&-\frac{1}{\sqrt{5}}(2\tau F_{H1}Q_{H1}-\frac{1}{2}(F_{H1}-\sqrt{5%
}F_{H2})Q_{H2}+\frac{1}{2}K_{H}(Q_{H1}^{2}+Q_{H2}^{2}).  \label{e17}
\end{eqnarray}

In this equation, $K_{H}$ is the harmonic force constant. The energy values
of $E(A)$ predicts the existence of saddle points.

The direct sum of $g$ and $h$ states consists of both $G$-type and $H$-type
nuclear coordinates. For this reason, computations are more complicated than
the $h$ state. In this case, there are three distortional coordinates that
are invariant under $D_{2}$. In Eqn.(16) nine dimensional matrix generators
which have been obtained by taking direct sum of five and four dimensional
representations are used to assign $Q_{i}$ values. It is found that $%
Q_{1}\rightarrow \sqrt{5}Q_{4}-\sqrt{3}Q_{5},\,Q_{2}\rightarrow
0,Q_{3}\rightarrow 0,\,\,Q_{4}\rightarrow Q_{4},\,Q_{5}\rightarrow
Q_{5},\,Q_{6}\rightarrow Q_{6},\,Q_{7}\rightarrow 0,\;Q_{8}\rightarrow \sqrt{%
5}Q_{6},\;Q_{9}\rightarrow 0$. In this basis, combination of eigenvalues of $%
B_{4}$ with harmonic potential gives us%
\begin{eqnarray}
E(A) &=&\pm \frac{1}{2\sqrt{10}}\left[ (3F_{H1}^{2}+F_{H2}^{2})(2\tau
^{2}Q_{H1}^{2}-5\tau Q_{H1}Q_{H2}+5Q_{H2}^{2})\right] ^{\frac{1}{2}}+\frac{3%
}{8}F_{G}Q_{G}+\frac{1}{2}K_{G}Q_{G}^{2}+\frac{1}{2}%
K_{H}(Q_{H1}^{2}+Q_{H2}^{2})  \notag \\
E(B_{1}) &=&\frac{1}{4\sqrt{5}}(F_{H1}+F_{H2})(4\tau
Q_{H1}-5Q_{H2})+(F_{H1}-3F_{H2})\sqrt{5}Q_{H2}-\frac{1}{4}F_{G}Q_{G}+\frac{1%
}{2}K_{G}Q_{G}^{2}+\frac{1}{2}K_{H}(Q_{H1}^{2}+Q_{H2}^{2})  \notag \\
E(B_{2}) &=&\frac{1}{4\sqrt{5}}(F_{H2}-F_{H1})(4\tau
Q_{H1}-5Q_{H2})+(F_{H1}+3F_{H2})\sqrt{5}Q_{H2}-\frac{1}{4}F_{G}Q_{G}+\frac{1%
}{2}K_{G}Q_{G}^{2}+\frac{1}{2}K_{H}(Q_{H1}^{2}+Q_{H2}^{2})  \notag \\
E(B_{3}) &=&-\frac{1}{\sqrt{5}}(F_{H1}22Q_{H1}-\frac{1}{2}(F_{H1}-\sqrt{5}%
F_{H2})Q_{H2}-\frac{1}{4}F_{G}Q_{G}+\frac{1}{2}K_{G}Q_{G}^{2}+\frac{1}{2}%
K_{H}(Q_{H1}^{2}+Q_{H2}^{2}),  \label{e18}
\end{eqnarray}%
where $K_{G}$ in the Eqn.(18) is the harmonic force constant of $G$ mode. It
is obvious that the magnitude of splitting of energy values is increasing in
hg state compared to h state.

\subsubsection{4.1.2 $D_{3}$ TRANSITION}

In this case decomposition of $H$ representation of irreducible
representations of $D_{3}$ is $A_{1}+2E$ in $D_{3}$. Following the same
procedure given in 4.1.1, symmetry of $I$ group can be broken into $D_{3}$,
in the directions, $Q_{1}\rightarrow 0,\quad Q_{2}\rightarrow 0,\quad
Q_{3}\rightarrow 0,\quad Q_{4}\rightarrow 0,\quad Q_{5}\rightarrow Q_{5}$.
The computed energy values are given as%
\begin{eqnarray}
E(A_{1}) &=&-\frac{2}{3}F_{H1}Q_{H2}+\frac{1}{2}K_{H}Q_{H2}^{2}  \notag \\
E_{\mp }(E) &=&\frac{1}{6}F_{H1}Q_{H2}\mp \sqrt{\left( \frac{F_{H1}}{3}%
\right) ^{2}+\left( \frac{F_{H2}}{2}\right) ^{2}}Q_{H2}+\frac{1}{2}%
K_{H}Q_{H2}^{2}.  \label{e19}
\end{eqnarray}%
Transformation of nuclear coordinates for $H\otimes (h\oplus g)$ state gives
that $Q_{1}\rightarrow 0,\,Q_{2}\rightarrow 0,Q_{3}\rightarrow
0,\,\,Q_{4}\rightarrow 0,\,Q_{5}\rightarrow Q_{5},\,Q_{6}\rightarrow
Q_{6},\,Q_{7}\rightarrow 0,\;Q_{8}\rightarrow 0,\;Q_{9}\rightarrow 0$. The
energy values consists of the two distortional coordinates, and therefore%
\begin{eqnarray}
E(A_{1}) &=&-\frac{2}{3}(F_{G}Q_{G}+F_{H1}Q_{H2})+\frac{1}{2}K_{G}Q_{G}^{2}+%
\frac{1}{2}K_{H}Q_{H2}^{2}  \notag \\
E_{\mp }(E) &=&\frac{1}{6}(F_{G}Q_{G}+F_{H1}Q_{H2})\mp \left[ \left( \frac{1%
}{3}F_{H1}Q_{H2}-\frac{5}{12}F_{G}Q_{G}\right) ^{2}+\frac{1}{4}%
F_{H2}^{2}Q_{H2}^{2}\right] ^{\frac{1}{2}}+\frac{1}{2}K_{G}Q_{G}^{2}+\frac{1%
}{2}K_{H}Q_{H2}^{2}\cdot  \label{e20}
\end{eqnarray}%
These energy values are in agreement with the energy values given in [1].

\subsubsection{4.1.3. D5 TRANSITION}

In $D_{5}$ the quintet irreducible representation state reduces to $%
A_{1}+E_{1}+E_{2}$. Applying the same procedure as in the previous section, $%
Q_{i}$ values are found as $Q_{1}\rightarrow -\frac{Q_{4}}{\tau ^{2}},\quad
Q_{2}\rightarrow 0,\quad Q_{3}\rightarrow 0,\quad Q_{4}\rightarrow 0,\quad
Q_{5}\rightarrow \frac{\sqrt{3}}{2\tau }Q_{4}$. The corresponding energies
are obtained as%
\begin{eqnarray}
E(A_{1}) &=&-\frac{2}{\sqrt{5}}F_{H2}Q_{H1}+\frac{1}{2}K_{H}Q_{H1}^{2}
\notag \\
E(E_{1}) &=&\frac{1}{2}(-F_{H1}+\frac{1}{\sqrt{5}}F_{H2})Q_{H1}+\frac{1}{2}%
K_{H}Q_{H1}^{2}  \notag \\
E(E_{2}) &=&\frac{1}{2}(F_{H1}+\frac{1}{\sqrt{5}}F_{H2})Q_{H1}+\frac{1}{2}%
K_{H}Q_{H1}^{2}\cdot  \label{e21}
\end{eqnarray}%
The calculations have been carried out for $H\otimes (h\oplus g)$ coupling
and same results have been obtained as expected.

Tetrahedral group ($T$) is also maximal little group of $G$ representation
of I group. In $H\otimes (h\oplus g)$ coupling some energy values have been
expected. Since ($T$ is not little group of $H$ representation. Thus in $%
H\otimes h$ coupling energy eigenvalues of $B_{2}$ is zero.) Decomposition
of $H$ state in $T$ group is $E+T$, and symmetry is broken according to the
directions $Q_{9}\rightarrow -\sqrt{\frac{5}{3}}Q_{6},\quad Q_{i}\rightarrow
0,\;\;(i=1,\ldots 8)$ . The corresponding eigenvalues of $B_{4}$are $%
F_{G}Q_{G}$ and $-\frac{2}{3}F_{G}Q_{G},$ for $E$ and $T$ respectively.

In general, minimal energy values for each little group can be obtained from
the given energy expressions.

\section{CONCLUSION}

In summary, we have shown how the symmetry breaking method is applied for
the determination of the potential energies of the $H\otimes (2h\oplus g)$
surface. In Ceulemans' paper [1], these energies were found by the method of
isostationary function and potential energies of $D_{3}$ and $D_{5}$ groups
were investigated. In our work, all maximal little groups of icosahedral
group are studied for the $H\otimes (h\oplus g)$ and $H\otimes h$ state.
Splitting of energy levels of icosahedral system due to distortion is
analysed and amazingly interesting that, there is a proper contribution on
the connection between our method and method of isostationary function. This
method can also be used for other distorted systems.

\textit{ACKNOWLEDGEMENTS}

Authors would like to thank the Scientific and Technical Research Council of
Turkey for its partial support.

\section{APPENDIX}

Generators of $I$, $D_{2},D_{3},D_{5}$ and $T$ groups for $G$and $H$
representations

Four and five dimensional irreducible matrix generators of $I$ and its
little groups can be generated from the matrices:

$\Omega ^{1},\Omega ^{2},\Gamma ^{1},\Gamma ^{2}\rightarrow I,\quad \Omega
^{3},\Omega ^{4},\Gamma ^{3},\Gamma ^{4}\rightarrow D_{5},\quad \Omega
^{2},\Omega ^{4},\Gamma ^{2},\Gamma ^{4}\rightarrow D_{3}$

$\Omega ^{1},\Omega ^{4},\Gamma ^{1},\Gamma ^{4}\rightarrow D_{2},\quad
\Omega ^{2},\Omega ^{5},\Gamma ^{2},\Gamma ^{5}\rightarrow T$%
\begin{equation*}
\Omega ^{1}=\frac{1}{3}\left(
\begin{array}{cccc}
-2 & 0 & \sqrt{5} & 0 \\
0 & 0 & 0 & 3 \\
\sqrt{5} & 0 & 2 & 0 \\
0 & 3 & 0 & 0%
\end{array}%
\right) ;\quad \Omega ^{2}=\frac{1}{2}\left(
\begin{array}{cccc}
2 & 0 & 0 & 0 \\
0 & -1 & \sqrt{3} & 0 \\
0 & -\sqrt{3} & -1 & 0 \\
0 & 0 & 0 & 2%
\end{array}%
\right)
\end{equation*}%
\begin{equation*}
\;\Omega ^{3}=\frac{1}{6}\left(
\begin{array}{cccc}
-4 & \sqrt{15} & -\sqrt{5} & 0 \\
-\sqrt{15} & -3 & \sqrt{3} & -3 \\
-\sqrt{5} & -\sqrt{3} & 1 & 3\sqrt{3} \\
0 & -3 & -3\sqrt{5} & 0%
\end{array}%
\right) ;\quad \Omega ^{4}=\left(
\begin{array}{cccc}
1 & 0 & 0 & 0 \\
0 & -1 & 0 & 0 \\
0 & 0 & 1 & 0 \\
0 & 0 & 0 & -1%
\end{array}%
\right)
\end{equation*}%
\begin{equation*}
\Omega ^{5}=\frac{1}{6\sqrt{3}}\left(
\begin{array}{cccc}
\sqrt{3} & -3\sqrt{5} & \sqrt{15} & -3\sqrt{5} \\
-3\sqrt{5} & 0 & -6 & -3\sqrt{3} \\
\sqrt{15} & -6 & -4\sqrt{3} & 3 \\
-3\sqrt{5} & -3\sqrt{3} & 3 & -3\sqrt{3}%
\end{array}%
\right) ;\Gamma ^{1}=\frac{1}{9}\left(
\begin{array}{ccccc}
7 & 0 & 0 & 2\sqrt{5} & -2\sqrt{3} \\
0 & 3\sqrt{5} & -6 & 0 & 0 \\
0 & -6 & -3\sqrt{5} & 0 & 0 \\
2\sqrt{5} & 0 & 0 & -1 & 2\sqrt{15} \\
-2\sqrt{3} & 0 & 0 & 2\sqrt{15} & 3%
\end{array}%
\right) \quad
\end{equation*}%
\begin{equation*}
\;\Gamma ^{2}=\frac{1}{2}\left(
\begin{array}{ccccc}
-1 & -\sqrt{3} & 0 & 0 & 0 \\
\sqrt{3} & -1 & 0 & 0 & 0 \\
0 & 0 & -1 & \sqrt{3} & 0 \\
0 & 0 & -\sqrt{3} & -1 & 0 \\
0 & 0 & 0 & 0 & 2%
\end{array}%
\right) ;\quad
\end{equation*}%
\begin{equation*}
\Gamma ^{3}=\frac{1}{36}\left(
\begin{array}{ccccc}
7-9\sqrt{5} & -7\sqrt{3}-3\sqrt{15} & 6\sqrt{3}-2\sqrt{15} & 18+2\sqrt{5} & 4%
\sqrt{3} \\
7\sqrt{3}+3\sqrt{15} & -21+3\sqrt{5} & -6-6\sqrt{5} & -6\sqrt{3}+2\sqrt{15}
& 12 \\
-6\sqrt{3}+2\sqrt{15} & -6-6\sqrt{5} & 3-3\sqrt{5} & -\sqrt{3}-3\sqrt{15} &
-12\sqrt{5} \\
18+2\sqrt{5} & 6\sqrt{3}-2\sqrt{15} & \sqrt{3}+3\sqrt{15} & -1+9\sqrt{5} & -4%
\sqrt{15} \\
4\sqrt{3} & -12 & 12\sqrt{5} & -4\sqrt{15} & 12%
\end{array}%
\right)
\end{equation*}%
\begin{equation*}
\quad \Gamma ^{4}=\left(
\begin{array}{ccccc}
1 & 0 & 0 & 0 & 0 \\
0 & -1 & 0 & 0 & 0 \\
0 & 0 & -1 & 0 & 0 \\
0 & 0 & 0 & 1 & 0 \\
0 & 0 & 0 & 0 & 1%
\end{array}%
\right)
\end{equation*}%
\begin{equation*}
\Gamma ^{5}=\frac{1}{12\sqrt{3}}\left(
\begin{array}{ccccc}
\frac{11}{\sqrt{3}}-\sqrt{15} & 7+\sqrt{5} & 10+2\sqrt{5} & -2\sqrt{\frac{5}{%
3}}+2\sqrt{3} & 2-6\sqrt{5} \\
7+\sqrt{5} & -\sqrt{3}+\sqrt{15} & -2\sqrt{3}-2\sqrt{15} & -14+2\sqrt{5} & -2%
\sqrt{3}-2\sqrt{15} \\
10+2\sqrt{5} & -2\sqrt{3}-2\sqrt{15} & 7\sqrt{3}-\sqrt{15} & -1-\sqrt{5} & -2%
\sqrt{3}+2\sqrt{15} \\
-2\sqrt{\frac{5}{3}}+2\sqrt{3} & -14+2\sqrt{5} & -1-\sqrt{5} & \frac{19}{%
\sqrt{3}}+\sqrt{15} & -6-2\sqrt{5} \\
2-6\sqrt{5} & -2\sqrt{3}-2\sqrt{15} & -2\sqrt{3}+2\sqrt{15} & -6-2\sqrt{5} &
-4\sqrt{3}%
\end{array}%
\right)
\end{equation*}

$\quad \quad \quad \quad $

$\quad \;\quad $

$\quad $

\textbf{REFERENCES}

1. A.Ceulemans, P.W. Fowler, J. Chem. Phys. 93, 1221 (1990).

2. A. Ceulemans, J. Chem. Phys. 87, 5374 (1987)

3. D. R.Pooler, J.Phys.C: Solid St.Phys. 13, 1029 (1980).

4. S.E. Apsel, C.C. Chancey, M.C.M O'Brien Phys. Rev. B 45, 5251 (1992).

5. N. Manini, E. Tosatti, A. Auerbach, Phys. Rev. B 49, 13008 (1994)

6. L. Michel, Rev.Mod.Phys. 52, 617 (1980).

7. M. Koca, R. Ko\c{c}, M. Al-Barwani, J.Phys.A: Math.Gen. 30, 2109 (1997).

8. R.Englman, The Jahn-Teller Effect in Molecules and Crystals (Wiley
--Interscience 1972).

9. P.L. Rios, N. Manini, E. Tosatti, Phys.Rev.B 54, 7157 (1996).

10. J. Ihm, Phys.Rev.B 49, 10726 (1994).

11. M.C.M. O'Brien, Phys.Rev.B 53, 3775 (1996).

12. C.P.Moate, J.L.Dunn,C.A.Bates and Y.M.Liu,Zeitschrift fur Physikalische
Chemie 200, 137 (1997)

13. P.W Fowler, A. Ceulemans, Molecular Physics 54, 767 (1985).

14. C.C. Chancey and M.C.M O'Brien, The Jahn-Teller Effect in C60 and Other
Icosahedral Complexes (Princeton University Press,1997).

\end{document}